# MULTI-PATHOLOGY CHEST X-RAY CLASSIFICATION WITH REJECTION MECHANISMS

Yehudit Aperstein[1], Amit Tzahar[1], Alon Gottlib[1], Tal Verber[1], Ravit Shagan Damti[1], Alexander Apartsin[2]

[1]Intelligent Systems, Afeka Academic College of Engineering, Tel Aviv 69988, Israel
[2]School of Computer Science, Faculty of Sciences, HIT-Holon Institute of Technology, Holon 58102, Israel

## ABSTRACT

Overconfidence in deep learning models poses a significant risk in high-stakes medical imaging tasks, particularly in multi-label classification of chest X-rays, where multiple co-occurring pathologies must be detected simultaneously. This study introduces an uncertainty-aware framework for chest X-ray diagnosis based on a DenseNet-121 backbone, enhanced with two selective prediction mechanisms: entropy-based rejection and confidence interval-based rejection. Both methods enable the model to abstain from uncertain predictions, improving reliability by deferring ambiguous cases to clinical experts. A quantile-based calibration procedure is employed to tune rejection thresholds using either global or class-specific strategies. Experiments conducted on three large public datasets (PadChest, NIH ChestX-ray14, and MIMIC-CXR) demonstrate that selective rejection improves the trade-off between diagnostic accuracy and coverage, with entropy-based rejection yielding the highest average AUC across all pathologies. These results support the integration of selective prediction into AI-assisted diagnostic workflows, providing a practical step toward safer, uncertainty-aware deployment of deep learning in clinical settings.

## 1. INTRODUCTION

Automating medical diagnosis with deep learning has shown great potential, particularly in medical imaging domains such as chest X-ray analysis. Convolutional neural networks, including architectures like DenseNet-121, have demonstrated strong performance in detecting a range of thoracic pathologies [1],[2]. However, successfully integrating such models into clinical workflows requires more than high classification accuracy. It demands robust mechanisms for managing uncertainty and ensuring patient safety.

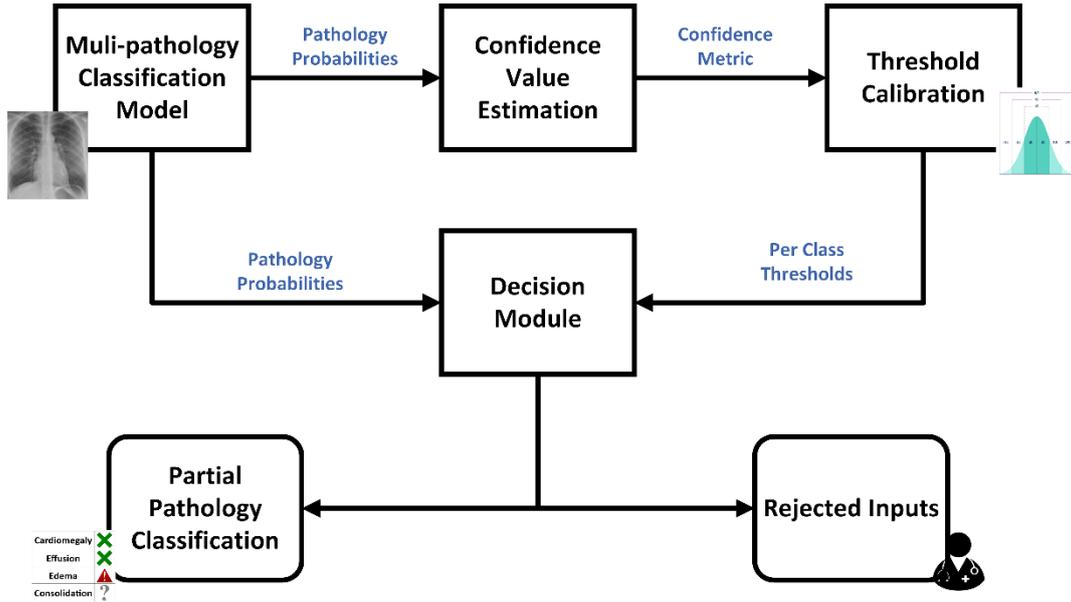

**Figure 1**: *Overview of the proposed selective chest X-ray classification framework*. The multi-pathology model produces per-class probability estimates, which are processed through a confidence estimation module. Based on calibrated thresholds, derived via entropy or confidence intervals, the decision module either accepts confident predictions (enabling partial classification) or rejects the input for expert review.

One of the most critical risks in this domain is the occurrence of false negatives, where existing pathologies go undetected. Such failures can delay treatment and lead to serious consequences for patient outcomes. The problem is compounded by the nature of medical imaging, where visual data alone may not be sufficient for a definitive diagnosis. Identifying a condition requires image-based interpretation, contextual clinical information, laboratory results, or expert judgment. These challenges are amplified in multi-label classification settings, where multiple overlapping or subtle findings must be simultaneously considered.

In high-stakes medical imaging, a model that always produces a prediction, even when uncertain, can pose significant risks. Rejection mechanisms provide a safeguard by allowing the model to abstain on ambiguous cases, deferring them to expert review. This capability is especially critical in multi-pathology chest X-ray classification, where subtle or overlapping findings can lead to overconfident errors. Integrating rejection directly into the diagnostic workflow, therefore, enhances reliability and patient safety beyond what conventional classifiers can provide.

To address these concerns, we propose a chest X-ray classification framework built on a DenseNet-121 backbone and augmented with two uncertainty-aware rejection mechanisms: (1) an entropy-based method, which measures per-class uncertainty using binary entropy, and (2) an interval-based method, which rejects predictions whose estimated confidence intervals encompass the decision threshold (typically 0.5). Both mechanisms allow the model to selectively abstain from uncertain predictions, deferring them to human experts for further evaluation, thus reducing the risk of overconfident misclassifications. By integrating both entropy-based and interval-based rejection into a unified framework, we enable complementary modes of abstention: entropy captures uncertainty from class probability distributions, while interval-based rejection enforces stricter decision-boundary separation. This dual design provides a richer spectrum of selective behavior than either method alone.

Figure 1 illustrates the core components of the proposed framework, highlighting the classification pipeline, the two rejection modules, and the calibration process that governs threshold selection. This visual overview complements the textual description by showing how uncertainty-aware decision-making is embedded into the model's workflow.

Crucially, we implement a calibration procedure using a held-out portion of the training data to tune rejection thresholds. We explore two strategies: shared thresholds, which apply uniformly across all pathology classes, and class-specific thresholds, which are independently optimized per class to account for differences in prevalence and model confidence. This calibration-driven approach enables flexible and interpretable control over the trade-off between diagnostic performance and rejection coverage, supporting safer and more trustworthy deployment of AI systems in real-world clinical settings.

This study makes the following contributions to the field of AI-assisted medical imaging:

(1) We propose an uncertainty-aware chest X-ray classification framework that integrates a standard DenseNet-121 backbone with two explicitly designed rejection mechanisms: an entropy-based approach that quantifies per-class predictive uncertainty and a confidence interval-based approach that rejects predictions overlapping the decision threshold. The novelty lies not in the backbone itself but in the systematic integration, calibration, and evaluation of these mechanisms for improving decision reliability in multi-pathology diagnosis.

(2) We design a quantile-based calibration procedure that supports both global and class-specific thresholds, enabling fine-grained control over the trade-off between diagnostic accuracy and rejection coverage.

(3) We evaluate the framework in both intra-source and inter-source settings across four major chest X-ray datasets, demonstrating improved robustness under domain shift and enhanced safety through uncertainty-aware abstention.

This study provides the first systematic evaluation of entropy-based and interval-based rejection mechanisms within the same framework, showing how each contributes to enhancing reliability in multi-pathology chest X-ray classification.

The remainder of this paper is organized as follows: Section 2 reviews related work on deep learning for chest X-ray diagnosis and selective classification. Section 3 describes the proposed methodology, including dataset construction, model design, and rejection mechanisms. Section 4 presents experimental results and comparative analysis. Section 5 discusses the findings and their implications. Finally, Section 6 concludes the paper and outlines directions for future research.

## 2. Prior work

Recent advances in deep learning have significantly improved automated chest X-ray interpretation, particularly in multi-label pathology classification. However, deploying such models in clinical settings requires more than high predictive accuracy. Frequently, it also demands robustness to uncertainty, transparency in decision-making, and mechanisms for risk-aware abstention. This section reviews relevant literature in three areas central to our work: deep learning for chest X-ray diagnosis, rejection mechanisms in machine learning, and the application of selective prediction and uncertainty modeling in medical imaging contexts.

### 2.1 Deep Learning in Medical Imaging and Chest X-ray Diagnosis

Deep convolutional neural networks (CNNs) have become the method of choice for medical image analysis. Large public datasets and deep CNNs enable expert-level disease classification in chest radiography. For example, [3] introduced the ChestX-ray8 dataset (~112,000 frontal CXRs, 14 labels) and demonstrated that CNNs can perform multi-label classification and localization of common thoracic diseases. [4] trained a 121-layer DenseNet ("CheXNet") on ChestX-ray14 and reported radiologist-level pneumonia detection (higher F1 score than average radiologist). The CheXpert dataset, released by Irvin et al. [5], includes 224,316 chest X-rays with uncertainty labels and has been used to show that CNN models can reach radiologist-level performance on several pathologies. In a smaller-scale study, [6] used AlexNet and GoogLeNet to classify tuberculosis versus healthy lungs on 1,007 CXRs, achieving an AUC of approximately 0.99. Majkowska et al. [7] trained deep models to detect five common CXR findings (e.g., pneumothorax, mass) and found performance comparable to radiologists. More recent work explores novel architectures: for instance, [8] applied a self-evolving vision transformer with knowledge distillation on chest X-rays, improving diagnostic accuracy by leveraging unlabelled data. In summary, foundational studies (and surveys) confirm that deep CNNs trained on large CXR datasets achieve state-of-the-art performance in pathology classification, setting the stage for further innovations.

Cohen et al. [9] explore how balanced batch sampling across multiple chest X-ray datasets can improve the out-of-distribution (OoD) generalization performance of deep learning models for pathology classification. They focus on four common conditions: Cardiomegaly, Consolidation, Edema, and Effusion. The dataset comprises four publicly available datasets: ChestX-ray8 [10], CheXpert [5], MIMIC-CXR[11], and PadChest[12]. Their main contribution is demonstrating that constructing mini-batches by sampling equally from each dataset (balanced batching) leads to more robust generalization than conventional random sampling from merged datasets.

The authors use a DenseNet-121 model fine-tuned on ImageNet features and evaluate it using a leave-one-dataset-out strategy. In this work, we augment the DenseNet-121 model with a rejection mechanism and assess the resulting system on the same four datasets.

## 2.2 Rejection Mechanisms in Machine Learning

Selective classification (the "reject option") allows a model to abstain on uncertain inputs to reduce errors. Early theory by Chow [13] formalized the error–reject trade-off. Building on this, Cortes et al. [14] developed formal frameworks for learning with a reject option, including boosting and empirical risk bounds. More recently, methods for deep nets have been proposed. Geifman and El-Yaniv [15] introduced a risk–coverage framework for DNNs: by thresholding the maximal SoftMax response, one can guarantee a user-specified error rate at high probability. Geifman and El-Yaniv [16] proposed *SelectiveNet*, a joint network+selector architecture that optimizes the accuracy–coverage trade-off during training. Another line of work calibrates or estimates uncertainty to drive abstention. Lakshminarayanan et al. [17] showed that *deep ensembles* (averaging multiple independently trained networks) provide high-quality uncertainty estimates without Bayesian complexity. Fisch et al. [18] focus on calibration: they train a separate selector to identify "uncertain" examples, aiming for predictions that are not only accurate but also well-calibrated (so that withheld examples have higher uncertainty). Cattelan and Silva [19] analysed post-hoc confidence for DNNs and found that SoftMax confidence can be "broken"; they propose a simple logit normalization (p-norm) that significantly improves confidence-based rejection. Rabanser et al. [20] take a different view: they track a model's training dynamics and reject inputs that continue to flip labels during training, achieving state-of-the-art error–coverage trade-offs without altering architecture or loss. In summary, classic methods (thresholding SoftMax or entropy, Chow's rule) and modern approaches (selective networks, deep ensembles, training-dynamics criteria) all provide mechanisms for a model to withhold predictions when unsure.

## 2.3 Rejection Methods in Medical Imaging and Chest X-rays Classification

Uncertainty-aware models can improve diagnostic safety by flagging ambiguous cases in medical contexts. Kompa et al. [21] review emphasizes that ML systems in healthcare should "say 'I do not know'" for high-uncertainty cases; they survey abstention methods and argue that uncertainty quantification (UQ) can make AI tools more reliable and trustworthy. In radiology specifically, Faghani et al. [22] outline emerging trends in UQ for medical imaging, noting that providing an uncertainty measure allows experts to re-evaluate doubtful cases. Concrete applications include chest X-rays: Whata et al. [23] developed a Bayesian CNN (using flipout layers) for multi-class CXR classification (COVID-19, pneumonia, normal) that explicitly quantifies predictive uncertainty. They further observe that standard CNNs often cannot express epistemic uncertainty, reinforcing the need for Bayesian or ensemble methods. Another example is Das et al. [24], who proposed *AnoMed*, a semi-supervised chest X-ray lesion detection framework. It uses confidence-guided pseudo-labelling: low-confidence (uncertain) pseudo-labels are filtered or refined so that only reliable labels are used in training, improving the localization of abnormalities. These works illustrate that rejection or uncertainty modelling is actively integrated into medical imaging pipelines to enhance decision support, particularly in critical tasks like diagnostic chest radiography.

## 3. Methodology

Our proposed framework for multi-label chest X-ray classification with selective prediction is illustrated in Figure 1. It integrates a deep convolutional neural network for pathology classification with two complementary rejection mechanisms that enable the model to abstain from uncertain predictions. In the subsections that follow, we detail the components of this framework, beginning with the dataset and data partitioning strategies, followed by the baseline model architecture, the design of the rejection mechanisms, and the threshold calibration procedure.

## 3.1 Dataset

The dataset comprises 122,830 chest X-ray images annotated with pathological labels. The classification task is formulated as multi-label, wherein each image can simultaneously exhibit one or more of four conditions: Cardiomegaly, Effusion, Edema, and Consolidation. In contrast to conventional single-label classification, this setup accounts for the co-occurrence of multiple pathologies within a single image. To improve generalizability and account for variability in imaging protocols and patient populations, the dataset aggregates samples from multiple publicly available sources. Each image is labeled at the image level (i.e., no spatial annotations), and the labels are not mutually exclusive. The model produces a confidence score for each image, which is subsequently utilized by the rejection mechanism to determine prediction reliability.

| Category | Positive Count | Percentage (%) |
| --- | --- | --- |
| Cardiomegaly | 11919 | 9.7 |
| Effusion | 9353 | 7.6 |
| Edema | 1956 | 1.6 |
| Consolidation | 2650 | 2.1 |
| Multi-Pathology | 3869 | 3.1 |

**Table 1**: Class distribution for the four target pathologies in the dataset

The class distribution is summarized in Table 1. The dataset is imbalanced, with a higher prevalence of Cardiomegaly and Effusion compared to the rarer findings of Edema and Consolidation. In addition, a substantial number of images (3,869; 3.15%) contain multiple positive labels, emphasizing the importance of designing models capable of handling multi-pathology predictions.

## 3.2 Data Partition for Validation and Rejection Tuning

Our dataset comprises chest X-ray images aggregated from four publicly available sources: CheXpert, MIMIC-CXR, NIH ChestX-ray14, and PadChest. Each dataset is generated using slightly different imaging protocols, labelling criteria, and patient populations. This diversity introduces domain variability, which challenges model generalization and reliable uncertainty estimation. Representative examples of chest X-rays labeled with various thoracic conditions, including the four pathologies studied in this work, are shown in Figure 2. These illustrate the visual diversity and diagnostic ambiguity often present in clinical data.

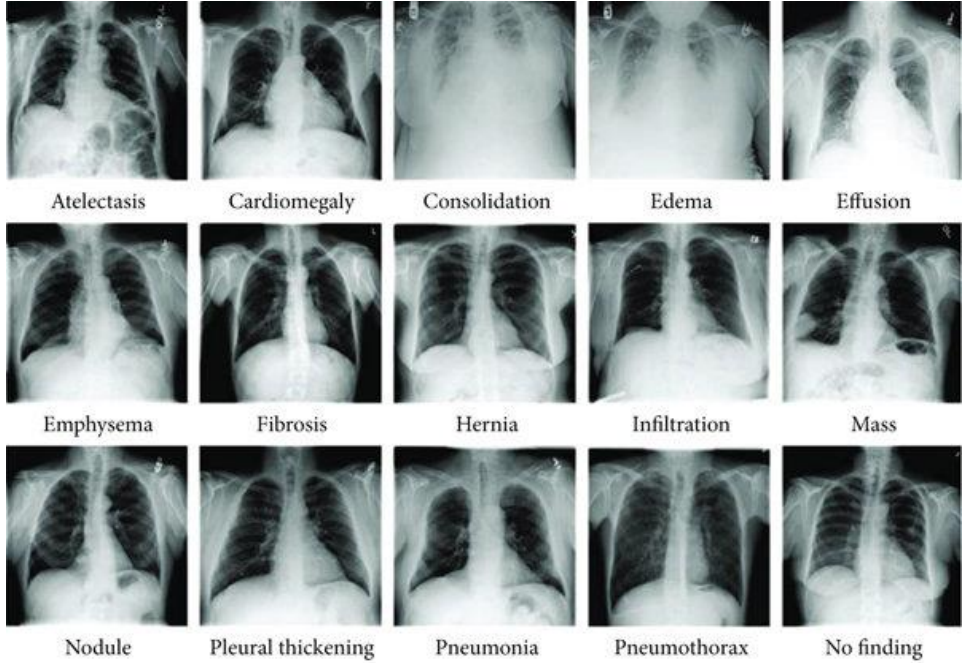

**Figure 2**: *Examples of chest X-ray images with their associated pathology labels.* The figure displays representative frontal chest radiographs corresponding to various thoracic conditions, including Cardiomegaly, Consolidation, Edema, and Effusion, which are among the four pathologies studied in this work.

We employed two data partitioning strategies to evaluate the model and tune the rejection mechanism: *Intra-Source Splitting* and *Inter-Source Splitting*. In Intra-Source Splitting, each dataset is split independently into 80% for training and 20% for validation, ensuring that both sets come from the same distribution. In Inter-Source Splitting, two or three complete datasets are used for training. In contrast, the remaining dataset(s) are held out entirely for validation, creating an out-of-distribution (OOD) test scenario. The intra-source approach supports stable tuning under controlled conditions, while the inter-source setup tests the model's robustness to domain shift and the rejection mechanism's effectiveness in unfamiliar settings.

## 3.3 BASELINE MODEL

DenseNet-121 [25] is a convolutional neural network architecture characterized by dense connectivity, wherein each layer receives, as input, the feature maps of all preceding layers and passes its output to all subsequent layers within the same dense block. This design enhances information flow, alleviates the vanishing gradient problem, and promotes feature reuse, allowing the network to be both deep and parameter efficient.

The full DenseNet-121 architecture consists of an initial convolution and pooling layer, followed by four dense blocks separated by three transition layers. Each dense block comprises multiple convolutional layers that follow a BN → ReLU → Conv(3×3) pattern. Within a block, each layer receives the concatenated outputs of all earlier layers in the same block, rather than just the output of the immediately preceding one. The growth rate $k$ determines how many new feature maps each layer contributes, thereby controlling the progressive expansion of feature dimensionality through the block. Transition layers between dense blocks perform downsampling and feature compression via convolution and pooling operations.

The baseline model employed for pathology classification in this study is based on the standard DenseNet-121 backbone architecture, which was initially pre-trained on large-scale chest X-ray datasets, including CheXpert and NIH ChestX-ray14, providing a diverse collection of labeled thoracic conditions across various patient populations and acquisition protocols [4], [5].

We employed a fine-tuning strategy that incorporates multi-domain balanced sampling to improve the model's ability to generalize across heterogeneous data sources. Rather than drawing training batches at random from the aggregated dataset, each mini-batch was constructed to include an equal number of samples from each domain. This enforced uniform representation during training prevents the model from overfitting to dominant sources and helps it learn domain-invariant features. Importantly, this strategy was applied regardless of class imbalance, focusing solely on balancing dataset origin. Prior work has shown that such domain-aware sampling improves robustness under domain shift and enhances out-of-distribution generalization in medical imaging classification tasks [9], [26]. The model performs multi-label classification by outputting a probability score for each target pathology: Cardiomegaly, Effusion, Edema, and Consolidation. The score reflects the likelihood that each condition is present in a given image. To evaluate model performance under class imbalance, we employ two widely used metrics: the area under the ROC curve (AUC), which measures ranking quality, and the F1 score, which balances precision and recall. These metrics are computed separately for each pathology and reported both before and after applying the rejection mechanisms. Hyperparameter tuning, including learning rate, batch size, and early stopping criteria, was conducted using a dedicated validation subset to optimize classification performance and avoid overfitting.

To better understand the behavior of the baseline model before applying any rejection mechanism, we examined the relationship between predictive uncertainty and classification errors across all datasets. Figure 3 presents representative results on the PadChest dataset. Incorrect predictions tend to occur at higher predictive entropy values, whereas correct predictions are concentrated at lower entropy values. This trend holds consistently across all target pathologies, including low-prevalence conditions such as Edema and Consolidation. These findings indicate that predictive uncertainty is a strong indicator of potential misclassification and motivate the integration of rejection mechanisms to improve decision reliability.

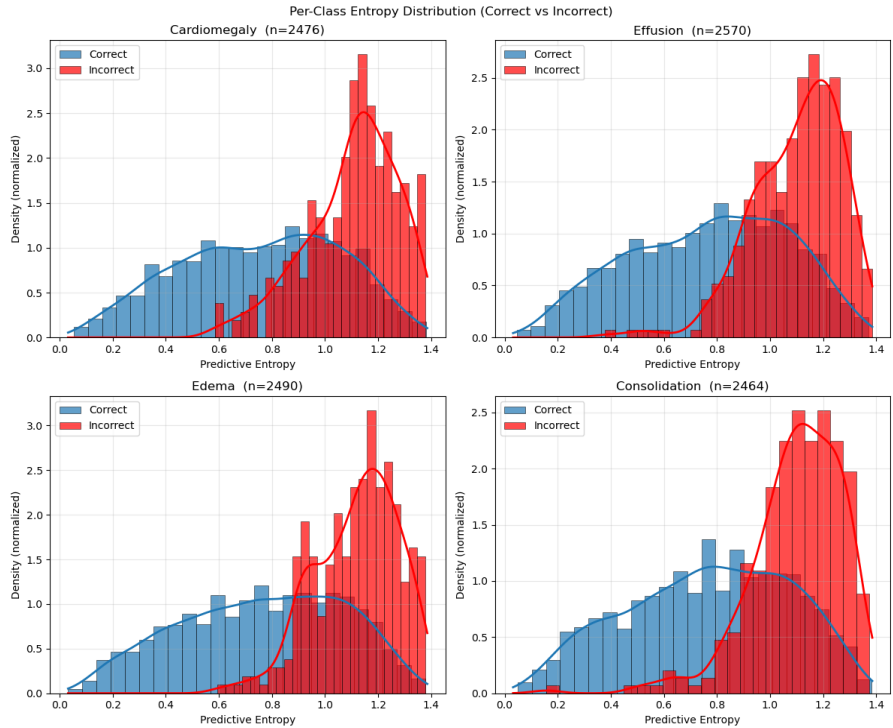

**Figure 3**: *Distribution of predictive entropy for correct and incorrect predictions across all pathology classes in the baseline model, shown here for the PadChest dataset.* Correct predictions cluster at low entropy values, while incorrect predictions are more prevalent at high entropy, highlighting the potential of uncertainty-based rejection to reduce misclassification risk.

## 3.4 REJECTION MECHANISM

The rejection mechanism in our framework plays a vital role in enhancing the safety and reliability of multi-label chest X-ray classification by enabling the model to abstain from making predictions when uncertainty is high. We explore two complementary approaches: entropy-based rejection and interval-based rejection, each using different confidence calibration criteria.

In the entropy-based method, the model produces a probability score $p_c \in [0,1]$ for each pathology class, and uncertainty is quantified using the binary entropy function, as defined in Equation 1:

$$H(p_c) = -p_c \log(p_c) - (1-p_c)\log(1-p_c) \quad (1).$$

Entropy is highest when the predicted probability is close to $p_c = 0.5$, that indicates the maximal uncertainty, and lowest when the model is confident (i.e., the predicted probability $p_c$ approaches 0 or 1. Instead of aggregating entropy across classes, we adopt a per-class evaluation rule: an image is accepted if at least one class prediction is considered confident, i.e., its entropy falls below a calibrated threshold $\tau_H$; it is rejected only if all predicted classes are deemed uncertain.

In the interval-based method, we evaluate whether the model's predicted probability for each class is statistically distinguishable from the decision boundary, typically set at $\theta = 0.5$. This is done by constructing a calibrated confidence interval $[p_c - \delta_c, p_c + \delta_c]$ for each class probability $p$. A class is marked as confident if its entire interval lies above or below the threshold, i.e. $p_c - \delta_c > \theta$ or $p_c - \delta_c < \theta$. As with the entropy-based method, an image is accepted if at least one class prediction meets the confidence criterion; otherwise, it is rejected.

For both methods, thresholds can be selected using either a shared (global) setting, where the same threshold is applied across all classes, or a class-specific setting, where thresholds are tuned independently per pathology. Calibration is performed using a held-out subset of the training data, and threshold selection aims to optimize the AUC on accepted predictions while keeping the rejection rate $R$ within a predefined bound $R \leq \varepsilon$, e.g., $\varepsilon = 10\%$. This dual-method, calibration-aware design enables nuanced control over model uncertainty and supports safe deployment in clinical environments.
.

## 3.5 THRESHOLD CALIBRATION

We adopt a quantile-based calibration strategy to determine effective thresholds for both entropy-based and interval-based rejection mechanisms. We aim to balance diagnostic performance with the rate of rejected predictions. The central objective is identifying thresholds that enable the model to confidently accept reliable predictions while deferring uncertain cases, enhancing overall system safety and clinical trust.

For entropy-based rejection, thresholds are calibrated using the distribution of entropy values computed from correctly classified samples in the training set. For interval-based rejection, the same quantile-based strategy is applied to the distribution of calibrated confidence margins, that is, the distance between the predicted class probability and the decision boundary (e.g., 0.5). In both cases, we explore a range of candidate percentiles (e.g., 75% to 95%) and, for each pathology, compute either a class-specific or global threshold corresponding to the selected quantile. This ensures that thresholds reflect uncertainty levels where the model has historically been accurate.

We evaluate each candidate to select the optimal percentile by plotting the resulting AUC on retained (non-rejected) predictions against the corresponding rejection rate. The final threshold maximizes AUC while constraining the rejection rate to a predefined limit (defaulting to 25%, though adjustable per clinical requirements). This quantile-based tuning procedure supports interpretable and flexible control over selective prediction behaviour in multi-label medical imaging tasks. Although we report results at fixed thresholds for consistency, both rejection mechanisms output continuous confidence scores, allowing thresholds to be tuned post hoc to balance diagnostic performance and coverage based on specific clinical requirements.

## 4. RESULTS

To evaluate the effectiveness of the entropy-based rejection mechanism, we designed a two-stage experiment involving a dedicated calibration set and an independent evaluation set, both drawn from the labelled chest X-ray dataset. This setup ensures that threshold tuning is performed without overfitting and that generalization performance is assessed fairly on unseen data.

We first applied the entropy-based rejection strategy, calibrating thresholds on the held-out calibration set using the quantile-based procedure described earlier. The evaluation set was then used to compute performance metrics, specifically, AUC and F1 score, before and after applying the rejection mechanism. These metrics were computed per pathology class across three source datasets: PadChest, NIH ChestX-ray14, and MIMIC-CXR.

Table 2 summarizes the results, including the calibrated entropy thresholds, baseline and post-rejection AUC and F1 scores, and the corresponding rejection rates.

| Dataset | Pathology Class | Entropy Threshold | AUC (baseline/with rejection) | F1 (baseline/ with rejection) | Rejection Rate |
|---|---|---|---|---|---|
| PadChest | Cardiomegaly | 0.58 | 77.2/**78.1** | 0.28/**0.30** | 15.53% |
| | Effusion | 0.62 | **89.0**/88.61 | **0.29**/0.26 | 35.98% |
| | Edema | 0.69 | 75.6/**78.78** | **0.64**/0.60 | 15.01% |
| | Consolidation | 0.63 | 80.7/**80.88** | 0.64/**0.67** | 25.10% |
| NIH ChestX-ray14 | Cardiomegaly | 0.58 | 68.7/**70.93** | 0.13/**0.18** | 25.23% |
| | Effusion | 0.64 | **80.9**/78.61 | 0.29/**0.36** | 45.73% |
| | Edema | 0.68 | 69.1/67.94 | 0.16/**0.19** | 29.41% |
| | Consolidation | 0.65 | **73.3**/68.04 | 0.14/**0.15** | 43.04% |
| MIMIC-CXR | Cardiomegaly | 0.66 | 82.7/**83.22** | 0.61/**0.63** | 2.91% |
| | Effusion | 0.69 | 90.7/**91.11** | 0.79/**0.81** | 3.03% |
| | Edema | 0.69 | 75.2/**75.36** | 0.33/**0.34** | 0.88% |
| | Consolidation | 0.69 | 82.9/**83.01** | 0.22/**0.23** | 1.69% |

**Table 2**: *Per-source classification results and rejection rates for entropy-based rejection. Metrics are reported both before and after rejection. Rejection thresholds were selected using a quantile-based calibration strategy on a held-out validation set.*

Across most pathology classes and datasets, the entropy-based rejection mechanism improves or maintains AUC, often at moderate rejection rates (e.g., 10–25%). For example, in the PadChest subset, the AUC improves for three out of four pathologies, and the F1 score improves or is preserved. In MIMIC-CXR, performance gains are achieved with minimal abstention (<3%), demonstrating the method's efficiency when trained and tested on similar distributions. Some drop in performance on the NIH dataset reflects sensitivity to domain shift, which is expected given its distinct data characteristics.

To benchmark the effectiveness of our entropy-based rejection strategy, we conducted a parallel experiment using the interval-based rejection method, following a procedure closely aligned with the entropy-based evaluation. The model was first trained on the same training set, and rejection thresholds were calibrated using a held-out calibration set. The final performance was assessed on an independent evaluation set. Table 3 presents a side-by-side comparison of the AUC scores achieved by the baseline model, entropy-based rejection, and interval-based rejection across the four target pathologies.

| Pathology Class | Baseline AUC | Interval-based Rejection AUC | Entropy-Based Rejection AUC |
|---|---|---|---|
| Cardiomegaly | 0.76 | 0.76 | **0.79** |
| Effusion | 0.79 | **0.87** | 0.86 |
| Edema | 0.73 | 0.73 | **0.79** |
| Consolidation | **0.87** | 0.81 | 0.77 |
| **Average over All Classes** | **0.79** | **0.81** | **0.83** |

**Table 3**: *Aggregated comparison of entropy-based and interval-based rejection methods*

Entropy-based rejection achieves the highest overall average AUC across the four pathology classes. However, interval-based rejection yields superior performance in specific cases, most notably for Consolidation and Effusion, where its conservative decision boundaries help reduce high-risk misclassifications. These differences likely reflect class-specific challenges. For instance, Edema and Consolidation are relatively rare in the dataset (1.6% and 2.2%, respectively), making them harder to learn and more susceptible to prediction uncertainty. This suggests that rejection decisions are concentrated in classes where the model is less confident due to limited training data or more ambiguous radiological features. Thus, integrating both rejection strategies helps balance sensitivity and specificity across pathologies with differing prevalence and diagnostic complexity.

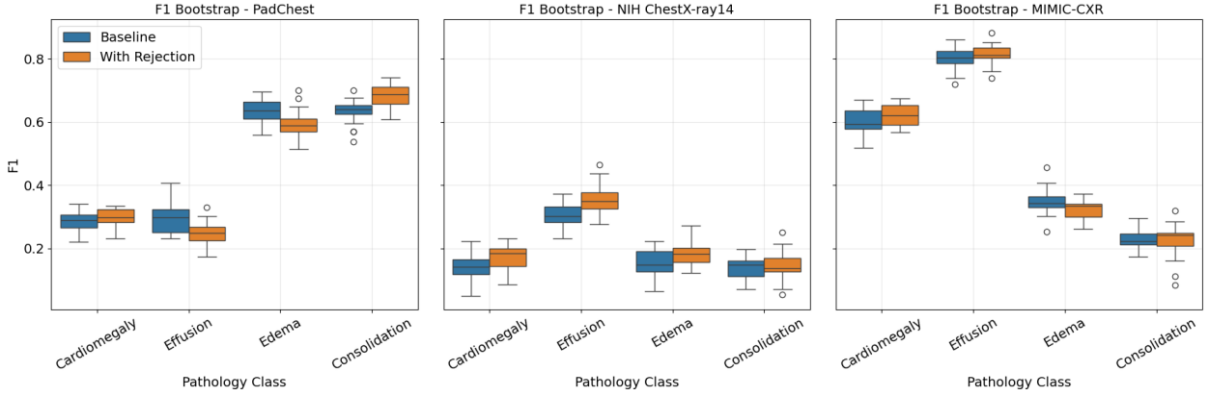

**Figure 4:** *Bootstrap-based statistical validation of F1 score changes before and after rejection.* Boxplots show F1 score distributions from 1,000 bootstrap resamples for each dataset and pathology. Blue: baseline; orange: with rejection.

To quantify the statistical significance of performance changes after applying the rejection mechanism, we conducted non-parametric bootstrap resampling (1,000 iterations) to derive F1 score distributions for each dataset–pathology pair (Figure 4). The analysis reveals distinct patterns across datasets: PadChest exhibits a mix of gains and declines across classes, NIH ChestX-ray14 shows broadly positive shifts, and MIMIC-CXR demonstrates small but consistent improvements. The degree of overlap between baseline and post-rejection distributions indicates that while some gains are statistically supported, others may reflect sampling variability. These findings confirm that the impact of rejection is both dataset- and class-dependent, underscoring the importance of uncertainty-aware tuning in diverse clinical contexts.

This comparison also serves as an ablation study, highlighting the individual contribution of each rejection mechanism. Entropy-based rejection achieves the highest overall average AUC across the four pathologies, indicating improved coverage–accuracy trade-offs. Interval-based rejection, while yielding a lower average, outperforms entropy on specific conditions such as Effusion and Consolidation, reflecting its stricter decision boundaries. Evaluating both within the same framework clarifies their complementary roles, with entropy supporting broader coverage and interval prioritizing conservative, high-certainty decisions.

These results demonstrate that both rejection mechanisms improve overall diagnostic performance, but their behavior differs. The entropy-based approach yields the highest average AUC across all four conditions, while the interval-based approach is more conservative and achieves the best AUC for Effusion. This highlights the complementary nature of the two methods: entropy-based rejection enables broader risk control and confidence filtering, whereas interval-based rejection offers a stricter, threshold-aware alternative suited to high-certainty scenarios.

## 5. CONCLUSIONS AND FUTURE DIRECTIONS

In this study, we presented a robust framework for chest X-ray diagnosis in a multi-label classification setting, addressing the clinical need for reliability in multiple, potentially overlapping pathologies. Our approach integrates a DenseNet-121 backbone with two selective prediction mechanisms: entropy-based rejection and interval-based rejection. We explored two calibration strategies for each method: global thresholds shared across classes and class-specific thresholds tailored to individual pathologies, using a dedicated calibration dataset to avoid overfitting and ensure generalizability.

The results demonstrate that rejection methods enable meaningful control over the trade-off between predictive confidence and coverage. By adjusting the rejection threshold, the system can effectively balance diagnostic performance (measured via AUC) against the fraction of predictions it chooses to defer. Entropy-based rejection offered a more flexible range of operations, while interval-based rejection provided a more conservative, risk-averse alternative. These findings support using uncertainty-aware selective prediction to enhance the safety and trustworthiness of AI-assisted diagnostic tools in clinical environments.

In this study, we have focused on analyzing the individual contribution of lightweight, interpretable rejection strategies for multi-pathology chest X-ray classification applied to a pretrained fixed backbone. The goal was to systematically investigate the effect of these strategies for individual and joint contribution to diagnostic reliability while isolating their effect from improvements that might be achieved due to more complex models with a built-in rejection mechanism or composite training objective.

These results also suggest that rejection behavior is influenced by pathology prevalence: rarer classes such as Edema and Consolidation, which account for less than 2.5% of the dataset, are more prone to uncertainty-driven abstention. By integrating entropy-based and interval-based rejection, the framework balances flexible and conservative risk control strategies, adapting to diagnostic challenges posed by class imbalance and multi-pathology complexity.

This study has several limitations. First, it relies solely on image data and does not incorporate complementary clinical information such as patient history or laboratory results, which could further enhance diagnostic performance. Second, although cross-source evaluations (Section 4.3) provide insight into model robustness across datasets, the study does not explicitly assess behavior under domain shift or out-of-distribution (OOD) conditions. Third, the proposed framework was evaluated on four thoracic pathologies; its applicability to broader diagnostic contexts remains to be validated. Future work should address these limitations by integrating multimodal inputs, formally evaluating OOD generalization, and extending the approach to additional disease classes and clinical use cases.

Future work can extend this framework by incorporating additional rejection techniques, such as Bayesian uncertainty estimation, Monte Carlo dropout, or model disagreement-based rejection. Moreover, using an ensemble of rejection strategies may offer complementary strengths and improve robustness. The baseline classification model can also be enhanced by evaluating alternative deep learning architectures such as EfficientNet, Vision Transformers, or hybrid convolutional-attention models, potentially leading to more accurate and well-calibrated diagnostic systems.

# DATA AVAILABILITY

All datasets used in this study are publicly available. CheXpert is available at https://stanfordmlgroup.github.io/competitions/chexpert/, MIMIC-CXR at https://physionet.org/content/mimic-cxr/2.0.0/, NIH ChestX-ray14 at https://nihcc.app.box.com/v/ChestXray-NIHCC, and PadChest at https://bimcv.cipf.es/bimcv-projects/padchest/.

# REFERENCES

[1] Sufian, M. A., Hamzi, W., Sharifi, T., Zaman, S., Alsadder, L., Lee, E., ... & Hamzi, B. (2024). AI-Driven Thoracic X-ray Diagnostics: Transformative Transfer Learning for Clinical Validation in Pulmonary Radiography. *Journal of Personalized Medicine*, *14*(8), 856.
[2] Arulananth, T. S., Prakash, S. W., Ayyasamy, R. K., Kavitha, V. P., Kuppusamy, P. G., & Chinnasamy, P. (2024). Classification of pediatric pneumonia using modified DenseNet-121 deep-learning model. *IEEE Access*.
[3] Wang, X., Peng, Y., Lu, L., Lu, Z., Bagheri, M., & Summers, R. M. (2017). ChestX-ray8: Hospital-scale chest X-ray database and benchmarks on weakly-supervised classification and localization of common thorax diseases. In *CVPR 2017* (pp. 3462–3471). IEEE.
[4] Rajpurkar, P., Irvin, J., Zhu, K., Yang, B., Mehta, H., Duan, T., ... & Ng, A. Y. (2017). CheXNet: Radiologist-level pneumonia detection on chest X-rays with deep learning. *arXiv preprint arXiv:1711.05225*.
[5] Irvin, J., Rajpurkar, P., Ko, M., Yu, Y., Ciurea-Ilcus, S., Chute, C., ... & Lungren, M. P. (2019). CheXpert: A large chest radiograph dataset with uncertainty labels and expert comparison. In *Proceedings of the AAAI Conference on Artificial Intelligence, 33*(1), 590–597.
[6] Lakhani, P., & Sundaram, B. (2017). Deep learning at chest radiography: Automated classification of pulmonary tuberculosis by using convolutional neural networks. *Radiology, 284*(2), 574–582.

[7] Majkowska, A. E., Mittal, S., Steiner, D. F., Reicher, J. J., McKinney, S. M., Duggan, G. E., ... & Lungren, M. P. (2020). Chest radiograph interpretation with deep learning models: Assessment with radiologist-adjudicated reference standards and population-adjusted evaluation. *Radiology, 294*(2), 421–431.

[8] Park, S., Kim, G., Oh, Y., Kim, J., Kim, S., Oh, H., ... & Rhee, K.-H. (2022). Self-evolving vision transformer for chest X-ray diagnosis through knowledge distillation. *Nature Communications, 13*, 3848.

[9] Cohen, J. P., Viviano, J., Morrison, P., Brooks, R., Hashir, M., & Bertrand, H. (2020). TorchXRayVision: A library of chest X-ray datasets and models.

[10] Xiaosong Wang, Yifan Peng, Le Lu, Zhiyong Lu, M. Bagheri, and R. Summers. Chestx-ray8: Hospital-scale chest x-ray database and benchmarks on weakly-supervised classification and localization of common thorax diseases. 2017 IEEE Conference on Computer Vision and Pattern Recognition (CVPR), pages 3462–3471, 2017.

[11] Johnson, A. E., Pollard, T. J., Greenbaum, N. R., Lungren, M. P., Deng, C. Y., Peng, Y., ... & Horng, S. (2019). MIMIC-CXR-JPG, a large publicly available database of labeled chest radiographs. *arXiv preprint arXiv:1901.07042*.

[12] Bustos, A., Pertusa, A., Salinas, J. M., Y De La Iglesia-Vaya, M. (2020). Padchest: A large chest x-ray image dataset with multi-label annotated reports. Medical image analysis, 66, 101797.

[13] Chow, C. (1970). On optimum recognition error and reject tradeoff. *IEEE Transactions on Information Theory*, *16*(1), 41-46.

[14] Cortes, C., DeSalvo, G., & Mohri, M. (2016b). Learning with rejection. *Journal of Machine Learning Research, 17*(216), 1–40.

[15] Geifman, Y., & El-Yaniv, R. (2017). Selective classification for deep neural networks. In D. Lee, M. Sugiyama, U. Luxburg, I. Guyon, & R. Garnett (Eds.), *Advances in Neural Information Processing Systems 30* (pp. 4878–4887). Curran Associates.

[16] Geifman, Y., & El-Yaniv, R. (2019). SelectiveNet: A deep neural network with an integrated reject option. In *Proceedings of ICML 2019* (pp. 2151–2160). PMLR.

[17] Lakshminarayanan, B., Pritzel, A., & Blundell, C. (2017). Simple and scalable predictive uncertainty estimation using deep ensembles. In I. Guyon et al. (Eds.), *Advances in Neural Information Processing Systems 30* (pp. 6405–6416). Curran Associates.

[18] Fisch, A., Jaakkola, T. S., & Barzilay, R. (2022). Calibrated selective classification. *Transactions on Machine Learning Research, 1*(2), 26.

[19] Cattelan, L., & Silva, R. (2024). How to fix a broken confidence estimator: A simple approach for post-hoc deep model calibration. *Proceedings of UAI 2024*.

[20] Rabanser, S., Thudi, A., Hamidieh, K., Dziedzic, A., & Papernot, N. (2022). Selective classification via neural network training dynamics. *Proceedings of UAI 2022*.

[21] Kompa, B., Snoek, J., & Beam, A. L. (2021). Second opinion needed: Communicating uncertainty in medical machine learning. *NPJ Digital Medicine, 4*, 4.

[22] Faghani, S., Moassefi, M., Rouzrokh, P., Khosravi, B., Baffour, F. I., Ringler, M. D., & Erickson, B. J. (2023). Quantifying uncertainty in deep learning of radiologic images. *Radiology, 308*(2) .

[23] Whata, A., Dibeco, K., Abiodun, O. I., Nakate, T., & Umerah, J. N. (2024). Uncertainty quantification in multi-class image classification using chest X-ray images of COVID-19 and pneumonia. *Frontiers in Artificial Intelligence, 7*, 1410841.

[24] Das, A., Gorade, V., Kumar, K., Chakraborty, S., Mahapatra, D., & Roy, S. (2024). Confidence-guided semi-supervised learning for generalized lesion localization in X-ray images. In *Med Image Comput Comput Assist Interv. MICCAI 2024* (LNCS 15001, pp. 242–252). Springer.

[25] Huang, G., Liu, Z., Van Der Maaten, L., & Weinberger, K. Q. (2017). Densely connected convolutional networks. In *Proceedings of the IEEE conference on computer vision and pattern recognition* (pp. 4700-4708).

[26] Seyyed-Kalantari, L., Liu, G., McDermott, M., Chen, I. Y., & Ghassemi, M. (2020). CheXclusion: Fairness gaps in deep chest X-ray classifiers. In *BIOCOMPUTING 2021: proceedings of the Pacific symposium* (pp. 232-243).